\documentclass[12pt]{iopart}
    \usepackage{epsfig}

\begin{document}
\title{Multisubband transport and magnetic deflection of Fermi electron trajectories in three terminal junctions and rings}
\author{M.R. Poniedzia{\l}ek and B. Szafran}

\address{Faculty of  Physics and Applied
Computer Science, AGH University of Science and Technology, al.
Mickiewicza 30, 30-059 Krak\'ow, Poland}

\date{\today}

\begin{abstract}
We study the electron transport in three terminal junctions and quantum rings looking for 
the classical deflection of electron trajectories in presence of intersubband scattering. We indicate that although the
Aharonov-Bohm oscillations and the Lorentz force effects co-exist in the low subband transport, for
higher Fermi energies a simultaneous observation of the both effects is difficult and calls for carefully formed
structures. In particular, in quantum rings with channels wider than the input lead the Lorentz force
is well resolved but the Aharonov-Bohm periodicity is lost in chaotic scattering events. In quantum rings 
with equal length of the channels and $T$-shaped junctions the Aharonov-Bohm oscillations are distinctly periodic but the Lorentz force effects are not well pronounced.
We find that systems with the wedge shaped junctions allow for observation of both the periodic Aharonov-Bohm oscillations
and the magnetic deflection. 
\end{abstract}
 \maketitle

\section{Introduction}

External magnetic field ($B$) applied perpendicular to the plane of confinement of a two-dimensional electron gas (2DEG) deflects the trajectories of electrons carrying the current flow. The deflection by the Lorentz force  -- responsible for the  Hall effect --
occurs also at the nanoscale and imprints classical features on the quantum transport phenomena.
The action of the Lorentz force was early detected in mesoscopic cross junctions \cite{Shepard}.
Cyclotron deflection of the electron trajectory was predicted
 for the coherent ballistic electron injection
through a quantum point contact (QPC)\cite{Usuki}. This deflection was later observed \cite{Crook}
by spatially resolved scanning gate microscopy \cite{review}. Semiclassical electron orbits \cite{js,uz} are observed in  magnetic focusing experiments \cite{Aidala} which detect peaks of conductance ($G$) between two QPCs for $B$ values which make the distance between the contacts equal to an integer multiple of the cyclotron diameter.  A theoretical description of the
magnetic injection in a T-shaped and wedge junctions
was given in Ref. \cite{Usuki2}. Recently, magnetic deflection in multiterminal quantum billiards was also  studied \cite{morfonios}.

For open quantum rings the classical deflection of the electron trajectories by the Lorentz force
competes with quantum interference effects \cite{szafranpeeters}, which are strong provided that the electron wave function
passes with equal amplitude through both the arms of the ring.
The preferential electron injection to one of the arms of the quantum ring by the Lorentz force reduces the visibility of the Aharonov-Bohm (AB) conductance oscillations in high $B$ \cite{szafranpeeters}.
Since low visibility of the AB oscillations may also result from decoherence, a conclusive experiment on the Lorentz force effect was proposed for a three-terminal quantum ring \cite{szafranpeetersepl}. According to the calculations
\cite{szafranpeetersepl}
the vanishing AB conductance oscillation at high $B$ should be accompanied by an imbalance of the electron transport probability to the left and right output leads. This behavior was indeed observed in subsequent conductance measurements  \cite{strambini} for a three terminal quantum ring.

The theoretical description of the Lorentz force for quantum rings \cite{szafranpeeters,szafranpeetersepl,szapo,koti} and the experiment \cite{strambini} studied the transport in the lowest subband of the transverse quantization.
However most of the experiments on the AB effects in quantum rings \cite{most} correspond to multisubband transport.
Wave packet dynamics in the second subband was described in Ref. \cite{chaves} which however considered a single-output-lead system very narrow channels
in which the Lorentz force effects are negligible. 
The purpose of the present paper is to describe the
effects of the Lorentz force in three terminal system when the stationary current flow at the Fermi level goes through several subbands.
A basic motivation for this study are the properties  of the asymptotic states within the channels (see below).
The charge density of the first subband is shifted to this edge of the channel which is preferred by the Lorentz force.
However, the charge density shift in the second subband is just opposite. Therefore, the Lorentz force effect in the multiband
transport needs to be clarified.

Below, we solve the scattering problem for several three-terminal systems using the wave function picture \cite{wfp} of the current
flow and evaluate the linear conductance in the Landauer-B\"uttiker approach \cite{lb}.
We discuss systems with a few types of junctions.  For quantum rings we analyze the Fourier transform of the transfer probabilities
searching for the peaks due to the AB effect and indicate that the the Lorentz force deflection leaves its signature in the low-frequency part of the transform.
We discuss the role of the width of the channels and the type of the junctions that allow the classical Lorentz force and the AB oscillations to co-exist.

\section{Theory}

    \begin{figure}[ht!]
     \centering
    \begin{tabular}{ll}
     (a) & (e) \\
     \hbox{\rotatebox{0}{
                    \includegraphics[bb=0 -40 450 384,  width=40mm] {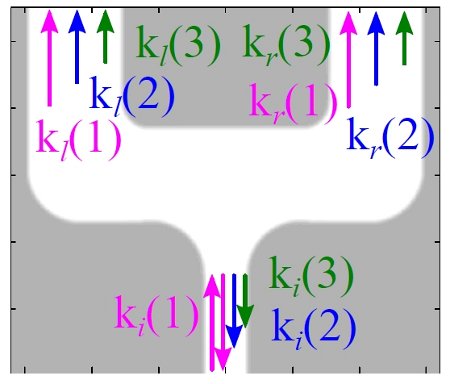}
                    }} &
    \hbox{\rotatebox{0}{
                    \includegraphics[bb=0 0 340 328,  width=40mm] {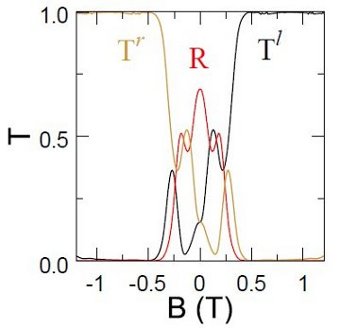}
                    }} \\
     (b) & (f) \\
    \hbox{\rotatebox{0}{
                    \includegraphics[bb=0 0 340 324,  width=40mm] {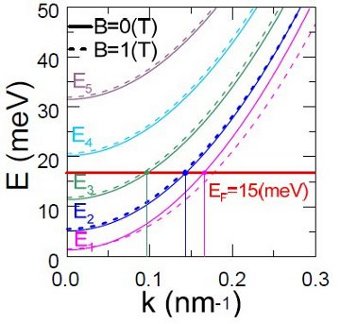}
                    }} &
    \hbox{\rotatebox{0}{
                    \includegraphics[bb=0 0 340 328,  width=40mm] {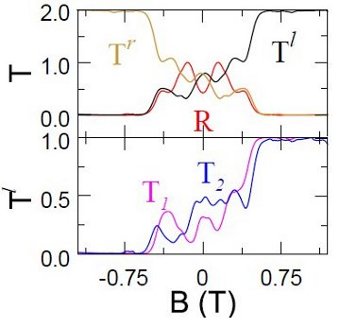}
                    }} \\
     (c) & (g) \\
    \hbox{\rotatebox{0}{
                    \includegraphics[bb=0 0 340 351,  width=40mm] {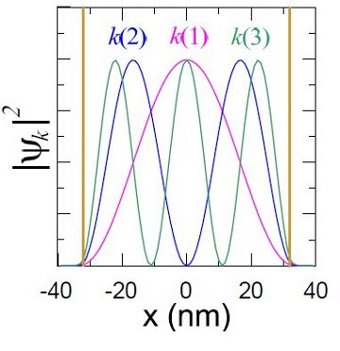}
                    }} &
    \hbox{\rotatebox{0}{
                    \includegraphics[bb=0 -25 340 329,  width=40mm] {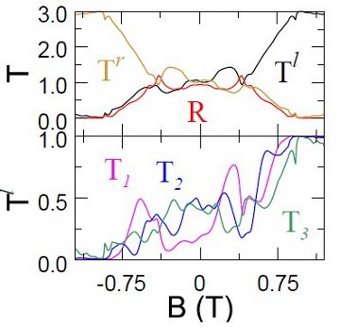}
                    }} \\
     (d) & (h) \\
    \hbox{\rotatebox{0}{
                    \includegraphics[bb=0 0 340 337,  width=40mm] {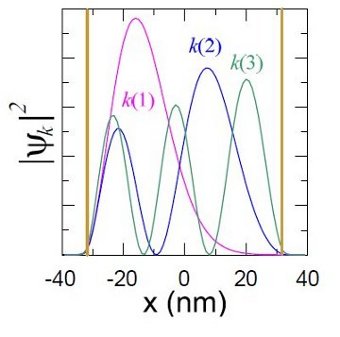}
                    }} &
    \hbox{\rotatebox{0}{
                    \includegraphics[bb=0 -20 340 319,  width=40mm] {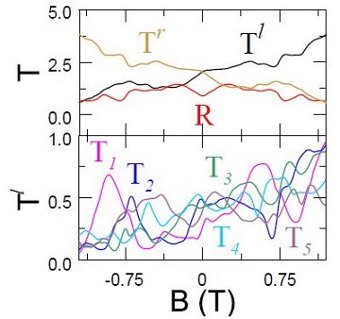}
                    }}

                    \end{tabular}

    \caption{
        (a) Geometry of a smooth and wide $T$ junction. In the solution of the scattering problem we assume that the electron
    comes to the system from the input lead (bottom of the figure) from a single subband $\mu$ with wave $k_(\mu)$. The electron
    may be backscattered to any subband of the input channel, or it is transferred to one of the subbands of the output leads.
    (b) Dispersion relation for a straight channel of width 64 nm and the symmetry axis at $x=0$ for $B=0$ (solid lines) and $B=1$ T (dashed lines). Horizontal line indicates the Fermi energy of 15 meV, which corresponds to three subbands appearing at the Fermi level -- the vertical lines indicate
     the Fermi wave vectors. Crossings of the Fermi energy level with the subband energies define the wave vectors $k_\mu$ participating in the transport. Charge densities across the channel for $E=15$ meV at $B=0$ (c) and $B=1$ T. (d)
     (e-f) Transfer probabilities to the left lead for electron incident from a given subband (left axis),
      the total transfer probability  to the left output lead (right axis, black lines), and backscattering probability $R$ (right axis, red lines).
      Calculations presented in (e), (f), (g) and (h) were performed for $E_F=3.5$, $6.5$, 15 and 36 meV, which corresponds to the single-, two-, three- and five- subband transport, respectively -- see (b).}
    \label{model}
    \end{figure}

In this paper we consider systems with three leads attached \cite{szafranpeetersepl,strambini} -- see for instance a junction displayed in Fig. \ref{model}(a). The input channel is the
one at the bottom of the Figure. We solve the scattering problem for a given Fermi energy $E_F$. For that purpose we consider the effective mass Schr\"odinger equation
\begin{equation}
\left[\frac{1}{2m^*}\left({\bf p}+e{\bf A}\right)^2 +V({\bf r})\right]\Psi({\bf r})=E_F\Psi({\bf r}),\label{hdft}
\end{equation}
with the assumption that the electron comes to the scattering region from a given subband $\mu$ of the input channel.
In Eq. (\ref{hdft}) $m^*$ and $V$ are the electron effective mass and the confinement potential, respectively.
We apply $m^*=0.067m_0$ for GaAs and for $V$ we take 0 inside the channels and 200 meV outside, which corresponds to GaAs/AlGaAs structure.

For strong confinement in the growth ($z$) direction that is present in 2DEG all the electrons occupy the same state of
the vertical quantization. The scattering problem can then be solved using a two-dimensional model, which we apply below.
In order to set the boundary conditions we need first to solve the Hamiltonian eigenequation in each of the leads. For
the Lorentz gauge ${\bf A}=(A_x,A_y,0)=(0,Bx,0)$, the electron eigenfunctions in the leads are separable into products of a transverse ($x$) and longitudinal ($y$) wave functions
$\psi_{k_\mu}=\exp(ik_\mu y)\Psi_{k_\mu}(x)$, where $k_{\mu}$ is the wave vector of $\mu$-th subband.
The eigenequation for the transverse wave function reads
\begin{equation}
\left[\frac{1}{2m^*}\left(-\hbar^2\frac{\partial^2}{\partial x^2}+(eBx+\hbar k)^2\right) +V_c(x)\right]\Psi(x)=E \Psi (x) \label{sb}.
\end{equation}
The dispersion relation calculated for a channel of width  64 nm is displayed in Fig. \ref{model}(b). A given Fermi energy fixes
the number of subbands participating in the transport along with the wave vectors $k_\mu$.
Application of the external magnetic field
breaks the parity symmetry of the transverse wave functions [cf. Fig. \ref{model}(c) and Fig. \ref{model}(d)].
For $B>0$ the term $2eBx\hbar k$ of the transverse Hamiltonian shifts the lowest subband solution to the left of the channel for the electron moving up the channel ($k>0$), in agreement with the orientation of the classical Lorentz force [see Fig. \ref{model}(d)].
The second subband wave function [see Fig. \ref{model}(d)] is shifted in the {\it opposite}  direction. This is because the wave function
of the second subband is {\it orthogonal} to the lowest-subband wave function with $k_2$.
The term of the Hamiltonian
which is responsible for the shifts is linear in $k$. Since $k_2<k_1$ the shift of the second subband wave function with $k_2$ to the right is weaker than the shift of the lowest subband function with $k_1$ to the left. The shift of the third subband density is
less pronounced due to a still  lower value of $k_3$.

For the electron incident from the input lead of the subband $\mu$ the wave function in the lower lead far away from
the scattering region -- outside the range of the evanescent \cite{es} modes -- has the form
\begin{equation}
\psi(x,y)=\exp(i k_\mu y)\Psi_{k_\mu}^\mu(x)+\sum_j b_j \exp(-i k_j y)\Psi_{-k_j}^j(x),
\end{equation}
where the sum goes over all subbands into which the electron may be backscattered.
In the left and right output lead the wave function consists of a superposition of states going out of the scattering
region
\begin{equation}
\psi(x,y)=\sum_j t^{l/r}_j \exp(i k^{l/r}_j y)\Psi_{k_j}^{l/r,j}(x).
\end{equation}

The backscattered ($b$) and the transferred ($t$) amplitudes are found via solution of the Schr\"odinger equation with
boundary conditions (3) and (4). The solution employs the finite difference approach.
The discretization of the equations and the self-consistent procedure for determination of the backscattered ($b$) and
transferred ($t$) amplitudes is given in detail in Ref. \cite{jasam}. The finite difference solution naturally
accounts for both appearance of evanescent modes in the Fermi level wave function within the scattering region as well
as for multisubband scattering.

The electron transfer probability from subband $\mu$ of the input channel to the subband $j$ of the output lead is then given by $T^{l/r}_{\mu,j}=\frac{\theta_j^{l/r}}{\theta_\mu}|t_j^{l/r}|^2$, where $\theta$'s are the fluxes of the probability density currents
integrated across the channels for the asymptotic Hamiltonian eigenstates.  In this paper we discuss the electron transfer probability to the left
or right output leads for the electron incident
from subband $\mu$, which is given by $T^{l/r}_\mu=\sum_{j} T^{l/r}_{\mu,j}$.
After summation over transfer probabilities over separate subbands we obtain transfer functions $T^{l/r}=\sum_\mu T^{l/r}_\mu$,
which are proportional to the conductance of both the output leads.
The conductance can be then evaluated as $G^{l/r}=\frac{e^2}{h}T^{l/r}$.

  \begin{figure}[ht!]

   \centering
    \includegraphics[scale=0.5,bb=0 0 600 336]{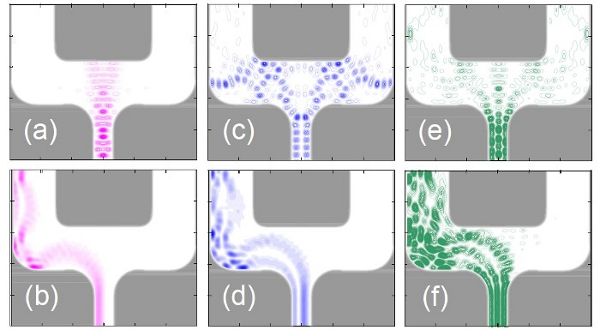}

    \caption{ The electron densities calculated
 for $E_F=15$ meV for $B=0$ (a,c,e) and $B=1$ T (b,d,f), for the electron incident from the lowest (a,b), second (c,d) and
 third (e,f) subbands.}
    \label{gest}
    \end{figure}

\section{Results}

Below, we consider systems with an input lead that is 64 nm wide, unless explicitly stated otherwise.

\subsection{Wider output channels and wedge junctions}
Let us first discuss a type of junction which corresponds to calculations of Ref. \cite{Usuki} -- see Fig. 1(a).
 The electrons are injected from the input lead of width 64 nm to a much wider perpendicular channel.
 The solutions of the scattering problem are plotted in Fig. 1(e-h).
For $B>>0$ is directed to the left output lead (see also Fig. 2) independent of the number of subbands participating in the charge transport and for each  subband from which the electron comes  to the junction.
Since for each incident subband at $B>0$ the electron tends to pass to the left output channel,
an overall slope of $T^l(B)$ increases with the number of occupied subbands.

The transfer probabilities for the quantum ring built in a similar manner [output channels and the ring wider than the input lead -- Fig. 3(a)] are shown in Fig. 3 (b,d,f) for one, two and three subbands appearing at the Fermi level.
The transfer and backscattering probabilities for the ring additionally contain a rapid oscillation which is due to the interference effects for the electron transfer amplitudes going through the left and right arms of the ring. However, the results -- positions of the rapid oscillation features -- do not exhibit any evident periodicity which is usually taken as the proof of the Aharonov-Bohm effect for the coherent current flow.

We performed the Fourier analysis of the transfer probability to the one of the output leads
\begin{equation} f(\nu)=\int dB \left( T^l(B)-\langle T^l\rangle \right) \exp(-2\pi i B\nu), \end{equation}
where $\langle T^l\rangle$ is the average transfer probability within the considered $B$ range.
The results for $F=|f|^2$ for the lowest subband transport are displayed in Fig. 3(c). The upper and
lower panels of Fig. 3(c) display low and high frequency ranges, respectively.
The low frequency part describes mostly the deviation of $T^l(B)$ off its average value which is strong
when the Lorentz force governs the electron flow.
The high frequency parts of the plot cover the region where the peaks due to the AB oscillation should be
expected to appear.

For the system of Fig. 3 the Fourier transform is entirely dominated by the low frequency part which results
from the distinct growth of $T^l$ from 0 to 1 within the studied range of $B$. This growth is exclusively due to the Lorentz force
which dominates the $T(B)$ characteristics.  The higher frequency part visible only under a close zoom  displays a peak near $\nu=8/$T, which corresponds to the magnetic field period of 0.125 T,
which in turn would correspond to the AB period for a ring of radius 102 nm -- somewhat smaller than
the mean radius of the ring of 120 nm [see Fig. 3(a)].
For $|B|>0.75$ T, the $T$ oscillations that are due to the quantum interference eventually disappear [Fig. 3(b)],
and then the transport is totally governed by the magnetic injection.
Lower panel of Figs. 3(b,d,f) displays the amplitude of oscillations of $T^l$ and $T^r$ which
is calculated for selected $B$ values (points at the plots) as the difference of the maximal and minimal $T$ values
within the range $(B-\Delta B/2,B+\Delta B/2)$, where $\Delta B=0.091$ T is the period  of AB oscillations for
a strictly one-dimensional ring of radius $R=120$ nm, i.e. the value of the magnetic field for which a magnetic flux quantum $\Phi_0=\pi(\Delta B)R^2=h/e$ threads the ring.

In Fig. 3(d) and (f) we can see that for two and three subbands at the Fermi level the oscillations of the transfer probabilities pertain up to higher values of $B$. The Fourier transform of $T^l$ are dominated by the low-frequency
part due to the Lorentz force, and the higher frequency part does not exhibit any pronounced peak indicating
any distinct period.

The studied system -- with the ring and the output channels that are wider than the input lead -- allows the Lorentz force
to dominate the current flow. The increased width of the channels leaves the space for magnetic deflection of their trajectories.
On the other hand, a part of lateral spatial quantization energy for the electron that enters the structure from the thinner input lead to
the wider channels is transformed to the kinetic energy of progressive motion which enhances the elastic and intersubband scattering effects.
Thus the system starts to resemble a cavity rather than a one-dimensional ring and the former is notorious for its chaotic transport properties.

  \begin{figure}[ht!]
   \centering
    \begin{tabular}{l}
    (a) \\

    \hspace{10mm} \hbox{\rotatebox{0}{
                    \includegraphics[bb=40 0 225 221, clip=true, width=40mm] {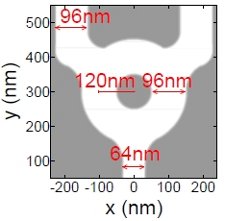}
                    }} \\
    (b)  \hspace{32mm} (c)\\
      \hbox{\rotatebox{0}{
                    \includegraphics[bb=0 0 510 365, clip=true, width=60mm] {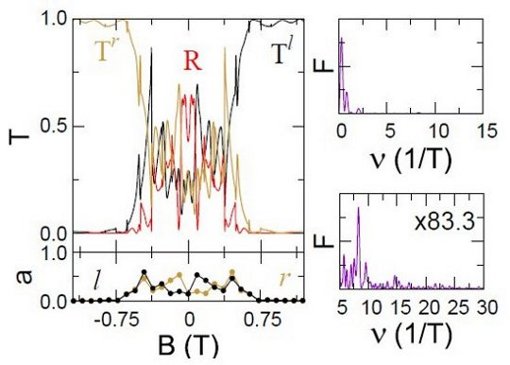}
					}} \\
		
    (d) \hspace{32mm} (e)\\
        \hbox{\rotatebox{0}{
                    \includegraphics[bb=0 0 510 365, clip=true, width=60mm] {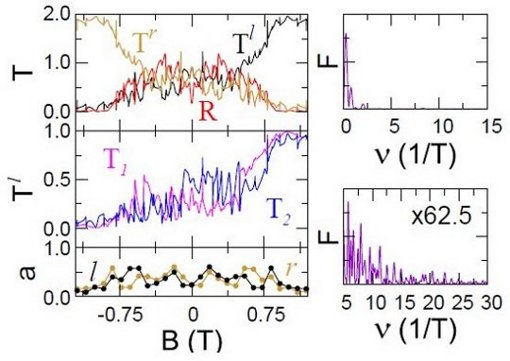}
                     }} \\
    (f) \hspace{32mm} (g)\\
      \hbox{\rotatebox{0}{
                    \includegraphics[bb=0 0 510 365, clip=true, width=60mm] {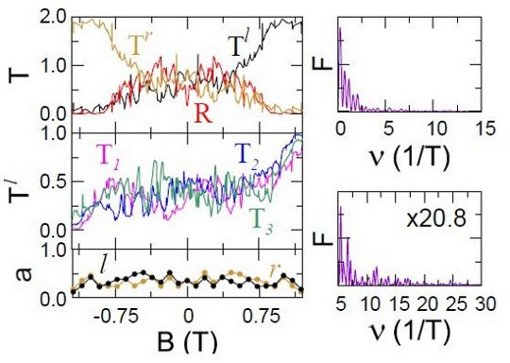}
                    }}

     \end{tabular}

    \caption{
(a) Studied structure with enlarged width of the channels within the ring and the output leads.
(b,d,f) transfer probabilities to the left and right output leads ($T^l$, $T^r$), the backscattering
probability ($R$), the transfer probabilities to the left output lead from separate subbands $T_\nu=T_\nu^l$
and the amplitudes of the oscillations $a$ for single, two--, and three subbands appearing at the Fermi level,
respectively ($E_F=3.5$ meV, 6.5 meV and 15 meV, respectively).  (c,e,g) show the Fourier transforms of $T^l$ in the low frequency part (upper plots) and the higher frequency part (lower plots). }
        \label{recTjun}
    \end{figure}

\subsection{Channels of equal length and wedge junctions}

Let us now consider the type of junction in which the input and output channels have similar width -- see Fig. 4(a) for the geometry
and Fig.4 (b,c,d) for the transfer probabilities at for two, three and five subband transport, respectively.
We notice in Fig.4  (c) and (d) that for higher subbands the guiding role of the Lorentz force is distinctly less effective
-- which is due to the properties of the channel eigenstates discussed above.
In consequence the overall growth of $T^l$ summed over the subbands from -1.2 T  to +1.2 T only weakly depends on the number of subbands participating
in the transport -- in contrast to the system studied in the precedent subsection.
Moreover, in Fig. 3(b,c) and (d) only $T_1^l$ probability is a monotonic function of the external magnetic field.
For the second subband only at higher $B$ the Lorentz force deflection of the trajectory wins with the shifts of the asymptotic channel wave function that occur due to the orthogonality conditions  [Fig. 1(d)].

The results for the quantum ring based on this type of junction -- with the channels that
are of similar width everywhere -- are displayed in Fig. 5. We find an imbalance of the electron transfer
due to the Lorentz force: $T^l>T^r$ at $B>0.75$ T
The oscillations of the transfer probabilities become more pronounced
and distinctly more regular than in the system with wider output channels [compare the amplitudes of Fig. 5(b,d,f) with Fig. 3(b,d,f)]. The Fourier spectra for
one and two subbands [Fig. 5 (c,e)] exhibit a well pronounced peak near $\nu=11$ / T, corresponding to the magnetic period of $0.091$ T -- for
the one dimensional ring of radius 120 nm -- equal to the average radius of the studied ring. A more complex structure is observed
in the Fourier transform for the three subband case -- still a presence of a maximum is evident -- in contrast to the previously discussed structure [Fig. 3(g)].

    \begin{figure}[ht!]
    \centering
    \begin{tabular}{l}
     (a)\\ \hspace{1mm}
     \hbox{\rotatebox{0}{
                    \includegraphics[bb=0 0 312 292,scale=0.5] {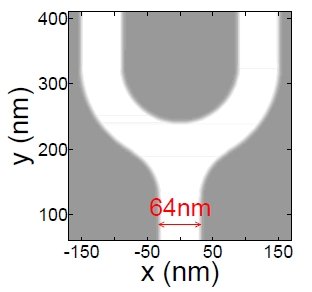}
                    }}  \\
     (b)\\
    \hbox{\rotatebox{0}{
                    \includegraphics[bb=0 0 320 308, clip=true, width=60mm] {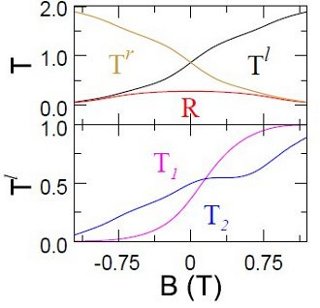}
                    }} \\
     (c)\\
    \hbox{\rotatebox{0}{
                    \includegraphics[bb=0 0 320 308, clip=true, width=60mm] {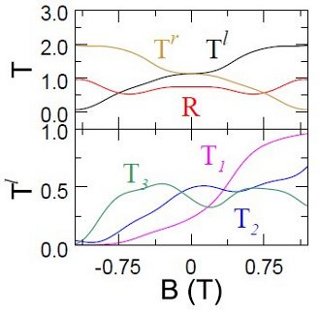}
                    }}\\

     (d)\\
    \hbox{\rotatebox{0}{
                    \includegraphics[bb=0 0 320 308, clip=true, width=60mm] {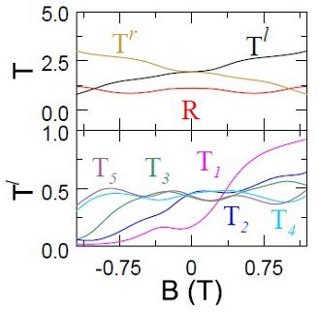}
                    }}

                    \end{tabular}

    \caption{Geometry of a wedge junction (a). (b,c,d) the upper panels show the transfer probabilities to the left $T^l$     and right $T^r$ output leads and the backscattering probability $R$ summed over the subbands the electron comes from. The lower panels in (b,d,d) show the subband-resolved transfer probability.  (b),  (c) and (d) correspond to 2, 3 and 5 subbands at the Fermi level, respectively ($E_F=6.5, 15, 34$ meV). }
    \label{smoTjun}
    \end{figure}

    \begin{figure}[ht!]
    \centering
    \begin{tabular}{l}
     (a) \\
     \hbox{\rotatebox{0}{
                    \includegraphics[bb=0 0 357 319, clip=true,width=50mm] {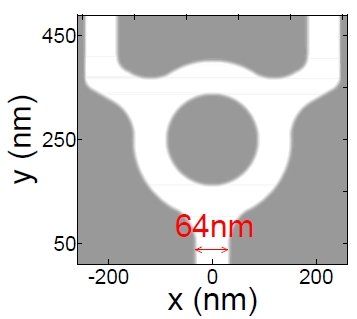}
                    }}  \\
     (b)  \hspace{32mm}   (c) \\
    \hbox{\rotatebox{0}{
                    \includegraphics[bb=0 0 510 373, clip=true, width=60mm] {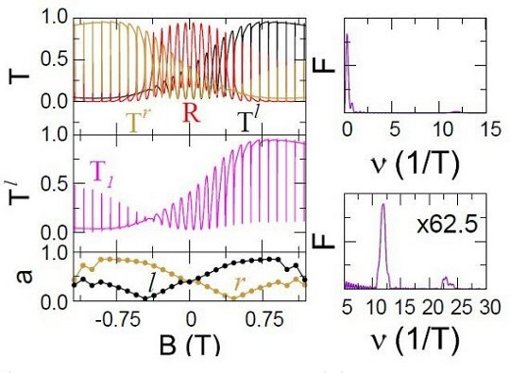}
                    }} \\
     (d) \hspace{32mm} (e) \\
    \hbox{\rotatebox{0}{
                    \includegraphics[bb=0 0 510 373, clip=true, width=60mm] {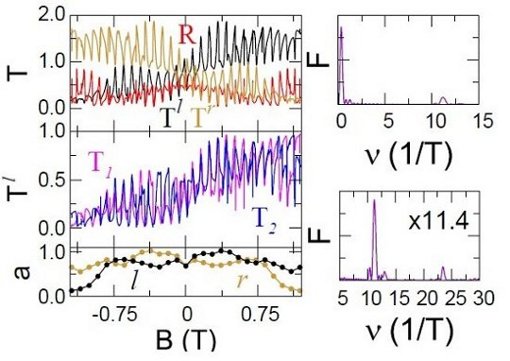}
                    }}\\
     (f) \hspace{32mm} (g) \\
    \hbox{\rotatebox{0}{
                    \includegraphics[bb=0 0 510 373, clip=true, width=60mm] {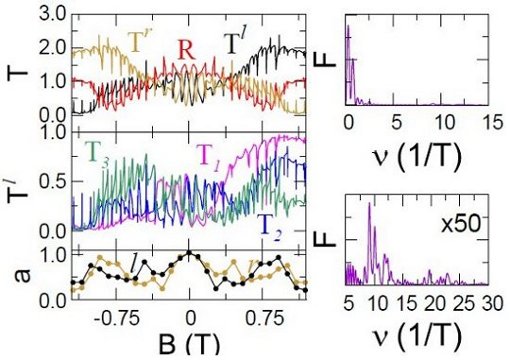}
                    }}

                    \end{tabular}

    \caption{Same as Fig. 3 only for a ring with equal length of the channels everywhere.}

    \label{ringRoundW64}
    \end{figure}

    \begin{figure}[ht!]
     \centering
    \begin{tabular}{l}
     (a)\\ \hspace{1mm}
      \hbox{\rotatebox{0}{
                    \includegraphics[bb=0 0 227 220, clip=true, width=40mm] {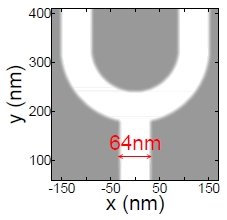}
                    }}  \\
     (b)\\
    \hbox{\rotatebox{0}{
                    \includegraphics[bb=0 0 320 302, clip=true, width=50mm] {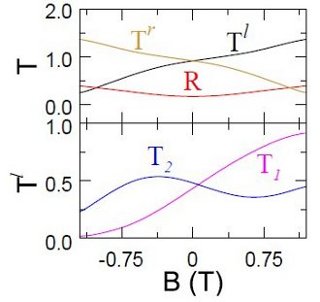}
                    }} \\
     (c)\\
    \hbox{\rotatebox{0}{
                    \includegraphics[bb=0 0 320 298, clip=true, width=50mm] {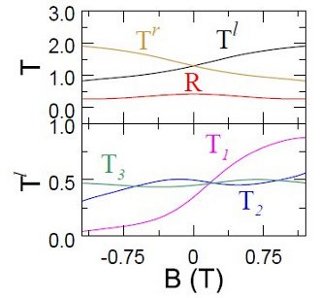}
                    }}\\

     (e)\\
    \hbox{\rotatebox{0}{
                    \includegraphics[bb=0 0 320 313, clip=true, width=50mm] {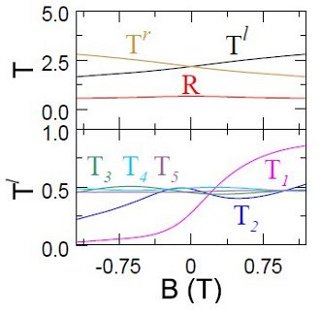}
                    }}

                    \end{tabular}

    \caption{Same as Fig. 3 only for $T$ junction replacing the wedge.}

    \label{recTjun1}
    \end{figure}

    \begin{figure}[ht!]
    \centering
    \begin{tabular}{l}
     (a) \\
     \hbox{\rotatebox{0}{
                    \includegraphics[bb=0 0 263 278, clip=true, width=40mm] {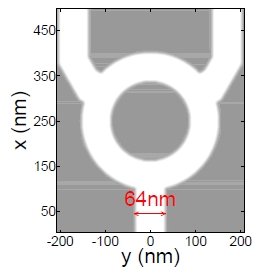}
                    }}  \\
     (b)  \hspace{32mm}   (c) \\
    \hbox{\rotatebox{0}{
                    \includegraphics[bb=0 0 510 351, clip=true, width=60mm] {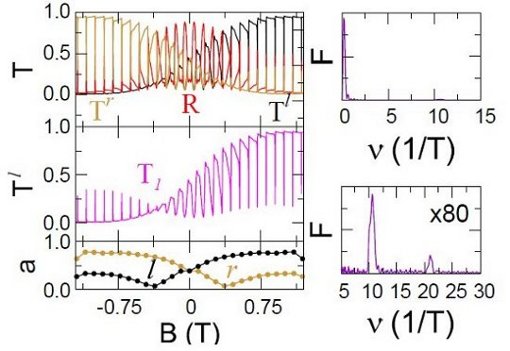}
                    }} \\
     (d) \hspace{32mm} (e) \\
    \hbox{\rotatebox{0}{
                    \includegraphics[bb=0 0 510 351, clip=true, width=60mm] {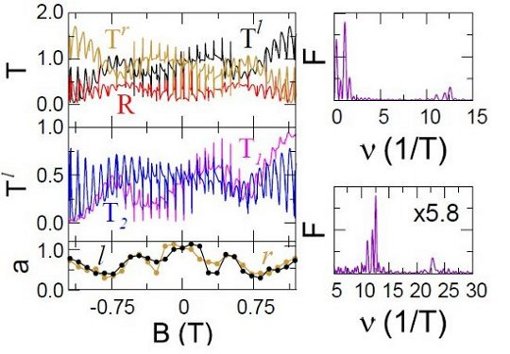}
                    }}\\

     (f) \hspace{32mm} (g) \\
    \hbox{\rotatebox{0}{
                    \includegraphics[bb=0 0 510 351, clip=true, width=60mm] {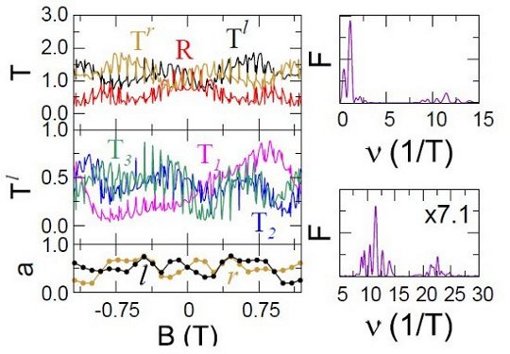}
                    }}

                    \end{tabular}

    \caption{Same as Fig. 5 only for $T$ junctions replacing the wedge junctions. }

    \label{ringRrectW64}
    \end{figure}

\subsection{Channels of equal length with T junctions}

The systems studied above contained a wedge shaped junctions. We find that the Lorentz force effects for these junctions are more evident than
for a simpler $T$ - shaped junctions. The results for a junction of this type are given in Fig. 6. As compared to Fig. 4 we notice that
$T_\mu^l$ for $\mu\ge 2$ becomes independent of $B$ within the studied range of the magnetic field. Moreover, the plot for $T_2^l$ indicates
that for the electron incident from the second subband at low magnetic field the current flows to the other output lead than the one which
is preferred by the Lorentz force. For the ring [Fig. 7] with this type of junctions we notice that the periodicity of the conductance oscillations
remains more or less similar to the one present for wedge-shaped junction [Fig. 5].
On the other hand, the imbalance of the
transfer probabilities to the two output channels is clearly introduced by the Lorentz force only in the lowest subband [Fig. 7(b)].

\subsection{Ring with thin channels}
The results for a ring with channels that are only 32 nm  wide are displayed in Fig. 8.
The results for the lowest subband channel - does not exhibit any trace of the magnetic deflection [Fig. 8(b)]-- note in particular the missing
low frequency part of the Fourier transport [Fig. 8(c)]. For the two-subband and three-subband transport we notice an imbalance of the conductance to the two output leads which is non-classical --  the current at $B>0$ is directed to the right output. This system -- close to a one-dimensional -- exhibits the clearest AB periodicity of the systems studied in this paper.

    \begin{figure}[ht!]
    \centering
    \begin{tabular}{l}
     (a) \\
    \hbox{\rotatebox{0}{
                    \includegraphics[bb=0 0 282 273, clip=true, width=40mm] {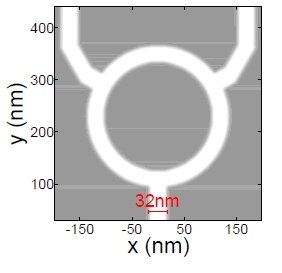}
                    }}  \\
     (b)  \hspace{32mm}   (c) \\
    \hbox{\rotatebox{0}{
                    \includegraphics[bb=0 0 510 351, clip=true, width=60mm] {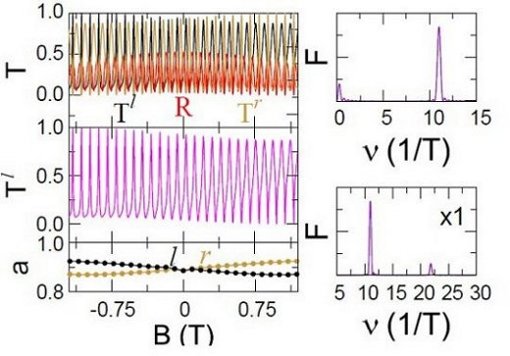}
                    }} \\
     (d) \hspace{32mm} (e) \\
    \hbox{\rotatebox{0}{
                    \includegraphics[bb=0 0 510 351, clip=true, width=60mm] {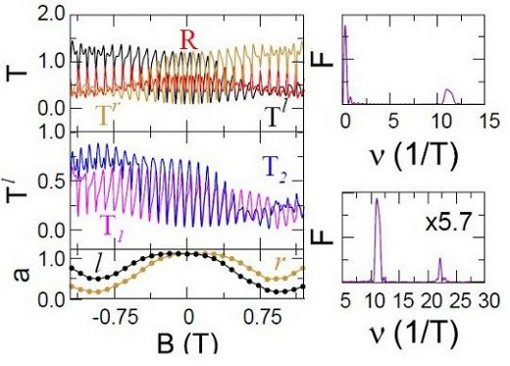}
                    }}\\
     (f) \hspace{32mm} (g) \\
    \hbox{\rotatebox{0}{
                    \includegraphics[bb=0 0 510 351, clip=true, width=60mm] {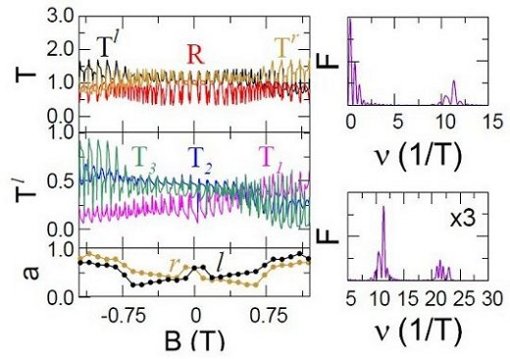}
                    }}

                     \end{tabular}
                      \caption{ Same as Fig. 7 only with channels of width equal to $32$ nm. (b-c,d-e,f-g) plots
                      were taken for Fermi energies $E_F=8,24,50$ meV, respectively}

    \label{ringRoundW64}
    \end{figure}

\section{Summary and conclusions}
We have studied the Lorentz force effects in three terminal junctions and quantum rings for both lowest subband
and multisubband transport conditions solving the stationary quantum scattering problem with a finite difference approach.

For the lowest subband transport the shift of the asymptotic wave function within the channels is consistent with
the orientation of the classical deflection. In consequence for transport in the lowest subband a distinct effect of the Lorentz force
and the pronounced Aharonov-Bohm oscillations coexist even for T type of junctions - in which the space for the electron deflection is quite limited.
Therefore, in quantum rings a clear Lorentz force effect and a pronounced AB oscillation appear in the lowest-subband transport as long
  as the channels are wide enough to allow for the magnetic deflection.
For higher Fermi energies corresponding to the multisubband transport usually more space is required for the classical electron deflection
than the one present in the T junction. The calculated transfer probabilities for multiband transport are distinctly Aharonov-Bohm periodic
but the effect of the Lorentz force is unclear if any.
One way to enhance the Lorentz force effects is to make the channels wider than the input lead. We demonstrated that this procedures
-- although successful for any subband -- leaves signatures of chaotic transport on the conductance which does not exhibit any periodicity.
Aharonov-Bohm oscillation and a distinct deflection of the electron trajectories can be obtained for wedge shaped junctions.

    {\bf Acknowledgements}
This work was performed
supported by the Polish Ministry of Science an Higher Education (MNiSW) within a research project N N202 103938 for 2010-2013.
 Calculations were  performed in
    ACK\---CY\-F\-RO\-NET\---AGH on the RackServer Zeus.


\begin{thebibliography}{00}

\bibitem{Shepard} K.L. Shepard, M.L. Roukes, and B.P. van der Gaas, Phys. Rev. B {\bf 46}, 9648 (1992).
        \bibitem{Usuki} T. Usuki, M. Takatsu, R.A. Kiehl, and N. Yokoyama  Phys.
Rev. B {\bf 50} 7615 (1994).
 \bibitem{Crook} R. Crook, C.G. Smith, M.Y. Simmons, and D.A. Ritchie, Phys. Rev. B {\bf 62}, 5174 (2000).
    \bibitem{review} H. Sellier, B. Hackens, M.G. Pala, F. Martins, S. Baltazar, X. Wallart, L. Desplanque, V. Bayot and S. Huant,
Sem. Sci. Tech. {\bf 26}, 064008 (2011).

    \bibitem{js} J. Schliemann, Phys. Rev. B {\bf 77}, 125303 (2008).
    \bibitem{uz} U. Z\"ulicke, J. Bolte and R. Winkler, New J. Phys. {\bf 9}, 355 (2007).
    \bibitem{Aidala} K. E. Aidala, R.E. Parott, T. Kramer, E.J. Heller, R.M. Westervelt, M.P. Hanson, and A.C. Gossard,
    Nature Physics {\bf 3}, 464 (2007).  
\bibitem{Usuki2} T. Usuki, M. Saito, M. Takatsu, R. A. Kiehl, and N. Yokoyama, Phys. Rev. B {\bf 52}, 8244 (1995).
\bibitem{morfonios} C. Morfonios, D. Buchholz, and P. Schmelcher, Phys. Rev. B {\bf 83}, 205316 (2011).
    \bibitem{szafranpeeters} B. Szafran and F.M. Peeters, Phys. Rev. B {\bf 72}, 165301 (2005).
    \bibitem{szafranpeetersepl}  B. Szafran and F.M. Peeters, Europhys. Lett. {\bf 70}, 810 (2005).
             \bibitem{strambini} E. Strambini, V. Piazza, G. Biasiol, L. Sorba, and F. Beltram, Phys. Rev. B {\bf 79}, 195443 (2009).
             \bibitem{szapo} M.R. Poniedzia{\l}ek and B. Szafran, J. Phys.: Condens. Matter {\bf 22}, 215801 (2010); J. Phys.: Condens. Matter {\bf 22} 468501 (2010).
             \bibitem{koti} V. Kotim\"aki and E. R\"as\"anen, Phys. Rev. B {\bf 81}, 245316 (2010). \bibitem{most}
             A. Fuhrer, S. L\"uscher, T. Ihn, T. Heinzel, K. Ensslin, W. Wegscheider, and M. Bichler, Nature (London) {\bf 413}, 822 (2001);
W. G. van der Wiel, Yu. V. Nazarov, S. De Franceschi, T. Fujisawa, J. M. Elzerman, E. W. G. M. Huizeling, S. Tarucha, and L. P. Kouwenhoven, Phys. Rev. B {\bf 67}, 033307 (2003); S. Pedersen, A. E. Hansen, A. Kristensen, C. B. Sorensen, and P. E. Lindelof, Phys. Rev. B {\bf 61}, 5457 (2000);
U. F. Keyser, C. F\"uhner, S. Borck, R. J. Haug, M. Bichler, G. Abstreiter, and W. Wegscheider, Phys. Rev. Lett. {\bf 90}, 196601 (2003);
A. M\"uhle, W. Wegscheider, and R.J. Haug, Appl. Phys. Lett. {\bf 91}, 133116 (2007); F. Martins, B. Hackens, M. G. Pala, T. Ouisse, H. Sellier, X.
Wallart, S. Bollaert, A. Cappy, J. Chevrier, V. Bayot, and S. Huant, Phys. Rev. Lett. {\bf 99}, 136807 (2007).
\bibitem{chaves} A. Chaves, G.A. Farias, F.M. Peeters, and B. Szafran, Phys. Rev. B {\bf 80} 125331 (2009). 
\bibitem{wfp} P. F\"oldi, O. K\`alman and M. G. Benedict, Phys. Rev. B {\bf 82}, 165322 (2010).
\bibitem{lb} M. B\"uttiker, Phys. Rev. Lett. {\bf 57}, 1761 (1986).
     \bibitem{es} P. F. Bagwell, Phys. Rev. B {\bf 41}, 10354 (1990).
     \bibitem{jasam} B. Szafran, Phys. Rev. B {\bf 84}, 075336 (2011).
\end{thebibliography}
\end{document}